\documentclass[10pt, conference, letterpaper]{IEEEtran}
\IEEEoverridecommandlockouts
\usepackage{cite}
\usepackage{amsmath,amssymb,amsfonts}
\usepackage{algorithmic}
\usepackage{graphicx}
\usepackage{textcomp}
\usepackage{xcolor}
\usepackage{bm}
\usepackage[normalem]{ulem}
\usepackage[colorlinks=true,citecolor=green,urlcolor=blue,bookmarks=false,hypertexnames=true]{hyperref} 
\usepackage{algorithm}
\usepackage{listings}
\usepackage{etoolbox}
\makeatletter
\AfterEndEnvironment{algorithm}{\let\@algcomment\relax}
\AtEndEnvironment{algorithm}{\kern2pt\hrule\relax\vskip3pt\@algcomment}
\let\@algcomment\relax
\newcommand\algcomment[1]{\def\@algcomment{\footnotesize#1}}
\renewcommand\fs@ruled{\def\@fs@cfont{\bfseries}\let\@fs@capt\floatc@ruled
  \def\@fs@pre{\hrule height.8pt depth0pt \kern2pt}%
  \def\@fs@post{}%
  \def\@fs@mid{\kern2pt\hrule\kern2pt}%
  \let\@fs@iftopcapt\iftrue}
\makeatother
\def\BibTeX{{\rm B\kern-.05em{\sc i\kern-.025em b}\kern-.08em
    T\kern-.1667em\lower.7ex\hbox{E}\kern-.125emX}}
\begin{document}

% \title{Paper Title*\\

\title{Mask Wearing Status Estimation with Smartwatches}
\author{\IEEEauthorblockN{Huina Meng$^{1*}$, Xilei Wu$^{1*}$, Xin Wang$^1$, Yuhan Fan$^2$, Jingang Shi$^1$, Han Ding$^1$, Fei Wang$^{1\#}$
}
\thanks{*equal contribution, \#corresponding author.}
\IEEEauthorblockA{\textit{1 Xi'an Jiaotong University} Xi'an Shaanxi, China, 710049 \\
\textit{2 Harbin Institute of Technology}, Harbin Heilongjiang, China, 150001 \\
\{menghuina, xlwuuu, xwang6\}@stu.xjtu.edu.cn, 1171910109@stu.hit.edu.cn, \{jingang, dinghanxjtu, feynmanw\}@xjtu.edu.cn}
}
% School of Software Engineering,
\maketitle

\begin{abstract}
We present MaskReminder, an automatic mask-wearing status estimation system based on smartwatches, to remind users who may be exposed to the COVID-19 virus transmission scenarios, to wear a mask. 
MaskReminder with the powerful MLP-Mixer deep learning model can effectively learn long-short range information from the inertial measurement unit readings, and can recognize the mask-related hand movements such as wearing a mask, lowering the metal strap of the mask, removing the strap from behind one side of the ears, etc. 
Extensive experiments on 20 volunteers and 8000+ data samples show that the average recognition accuracy is 89\%. Moreover, MaskReminder is capable to remind a user to wear with a success rate of 90\% even in the user-independent setting.
\textit{Code is publicly available at \url{https://github.com/aiotgroup/MaskReminder}}

\end{abstract} 

% such as those who are caring for cases of COVID-19, awaiting in queues for COVID-19 test, with symptoms suggestive of COVID-19, etc. 
% Wearing a mask is one of the simple yet critical methods to suppress the virus transmission in the current COVID-19 pandemic as the World Health Organization advised. 
% In this paper, 
% MaskReminder is designed to recognize the mask-related hand movements such as wearing a mask, lowering the metal strap of the mask, removing the strap from behind one side of the ears, etc. To make MaskReminder powerful, we apply the MLP-Mixer model that is with the strength of learning local and global representation across the temporal time-serial data. Experimental results from over 20 volunteers show that accuracy is 89\%, which is useful to remind the users to wear a mask. 

% \begin{IEEEkeywords}
% Smartwatches, mask wearing, deep learning, COVID-19
% \end{IEEEkeywords}

\section{Introduction}~\label{sec:introduction}
As World Health Organization~(WHO) advised the public, wearing a mask is one of the simple yet critical precautions to suppress transmission and save lives during the COVID-19 pandemic~\cite{whomask}. The purposes of wearing a mask are basically protecting healthy people from becoming infected, or preventing transmission from a wearer who is infected and may or may not have symptoms. Thus, many people should wear a mask in scenarios such as when they are caring for cases of COVID-19, awaiting in queues for COVID-19 tests, with suggestive symptoms of COVID-19, etc. Considering the importance of wearing masks, in this paper, we propose a smartwatch-based mask-wearing reminder, named MaskReminder, for those who are involved in the transmission scenarios mentioned but not limited to above.

MaskReminder is possible based on two clear observations. First, if we wear a mask, the wearing activity, comprised of a sequence of steps, makes it distinguishable from other activities such as walking, running, typing, etc. Second, if we have already worn a mask, we always tend to do some related activities, e.g., removing the strap from behind one side of ears shortly for food or breath, lowering the top metal strap to expose the mouth for breath, pinching the metal strap occasionally to mold the shape of the nose, etc. We utilize the inertial measurement unit~(IMU) of smartwatches, i.e., accelerators and gyroscopes, to characterize the user's hand movements, and enable MaskReminder to estimate the status of whether a user wears a mask or not. 

\begin{figure}[t]
    \centering
    \includegraphics[width=1\linewidth]{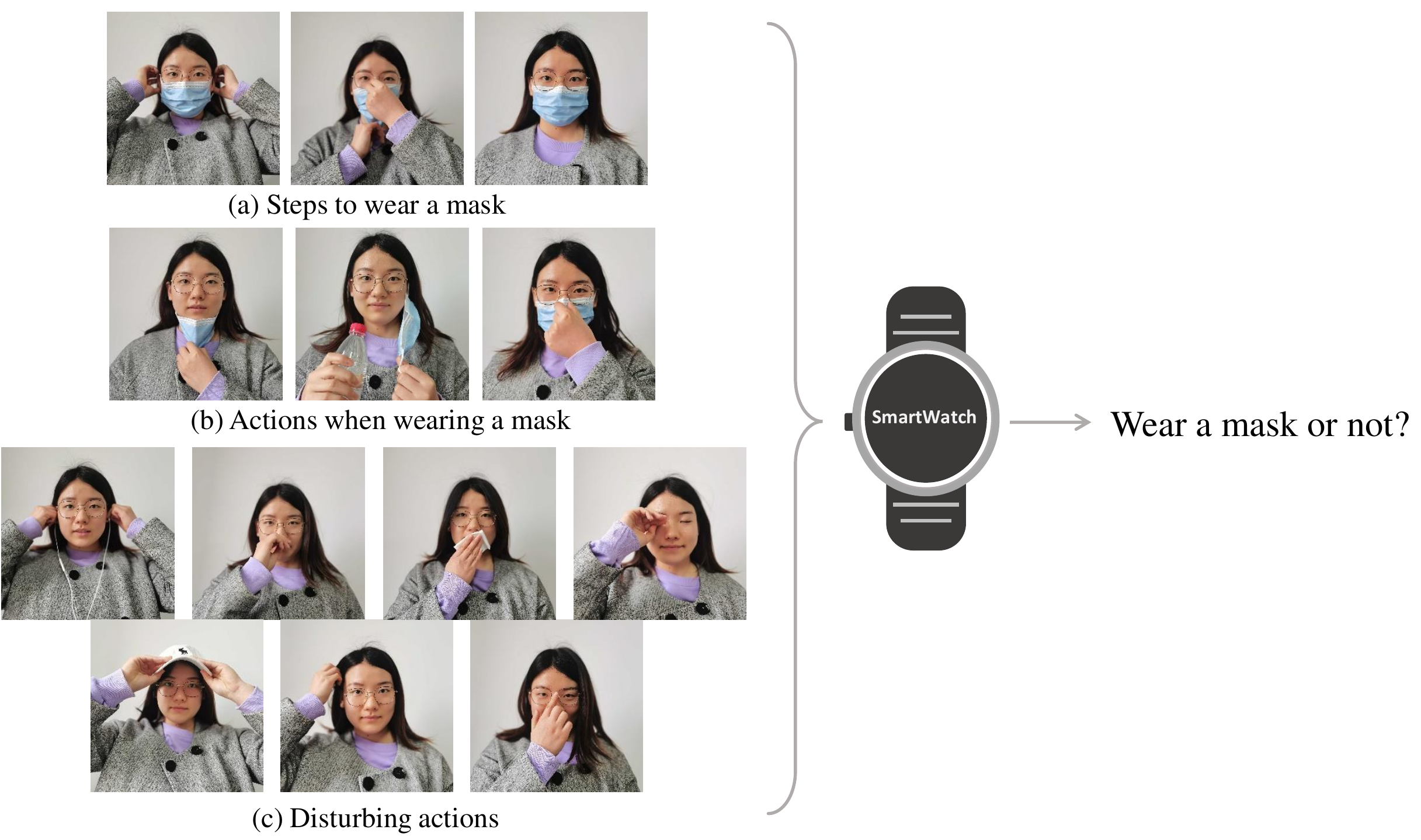}
    \setlength{\abovecaptionskip}{-10pt}
    \caption{MaskReminder, a smartwatch-based system, can estimate the mask-wearing status of users according to the mask-related activities such as wearing a mask, lowering the metal strap for breath, removing one strap behind the ears for drink, etc.}
    \label{fig:fig1}
\end{figure}

Broadly, characterizing the readings of IMU is the well-known task of time series recognition. The characteristics of time series include local/short-range information such as shapelet, saltation and trend, as well as global/long-range information such as extreme values, seasonal periods and shape. Thus, our first challenge is to extract informative and distinguishable representations from both local and global aspects of the accelerators and gyroscopes. 

% Existing approaches always pay imbalanced attention to local information and global information. (1) Support Vector Machine~(SVM). In~\cite{svm}, authors utilized global statics of time series, e.g., extreme values,  spectral, and phase domain features, and applied SVM to classify drivers' drowsiness.  (2) Dynamic Time Warping~(DTW), DTW aligns two series and computes the distance between them, which emphasizes the global shape when aligning. (3) Convolutional Neural Networks~(CNNs), The convolution operation has intrinsic locality bias due to the small size of convolution kernels. Though convolution layers are always stacked deeply to enlarge the receptive fields, the information interaction between far ranges only processes at high-level layers, much less than those in local ranges. 

% (4) Recurrent Neural Networks~(RNNs). For the computation on the inputs is conducted serially, RNNs are prone to extract representations from the nearest inputs.     
\begin{figure*}[t]
    \centering
    \includegraphics[width=1\linewidth]{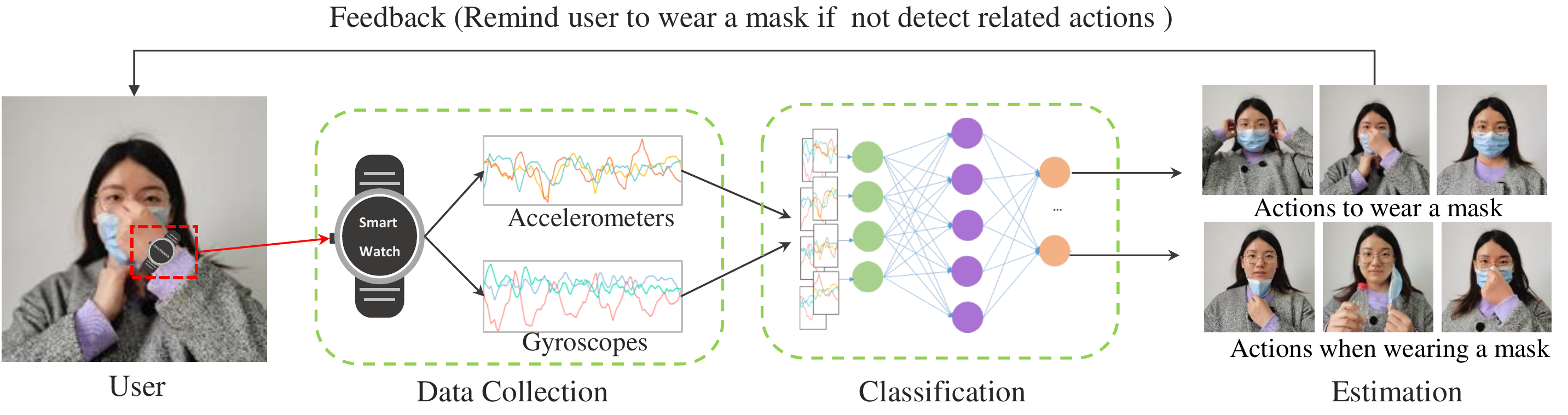}
    \setlength{\abovecaptionskip}{-10pt}
    \caption{System workflow. MaskReminder reads the recordings of accelerometers and gyroscopes of smartwatches continuously and conducts mask-wearing status estimation over the recordings. MaskReminder will keep silent if it detects the user has worn a mask in a recent period. Otherwise, it will pop up a notification to alert the user to wear a mask.}
    \label{fig:workflow}
\end{figure*}

\begin{figure}[t]
    \centering
    \includegraphics[width=1\linewidth]{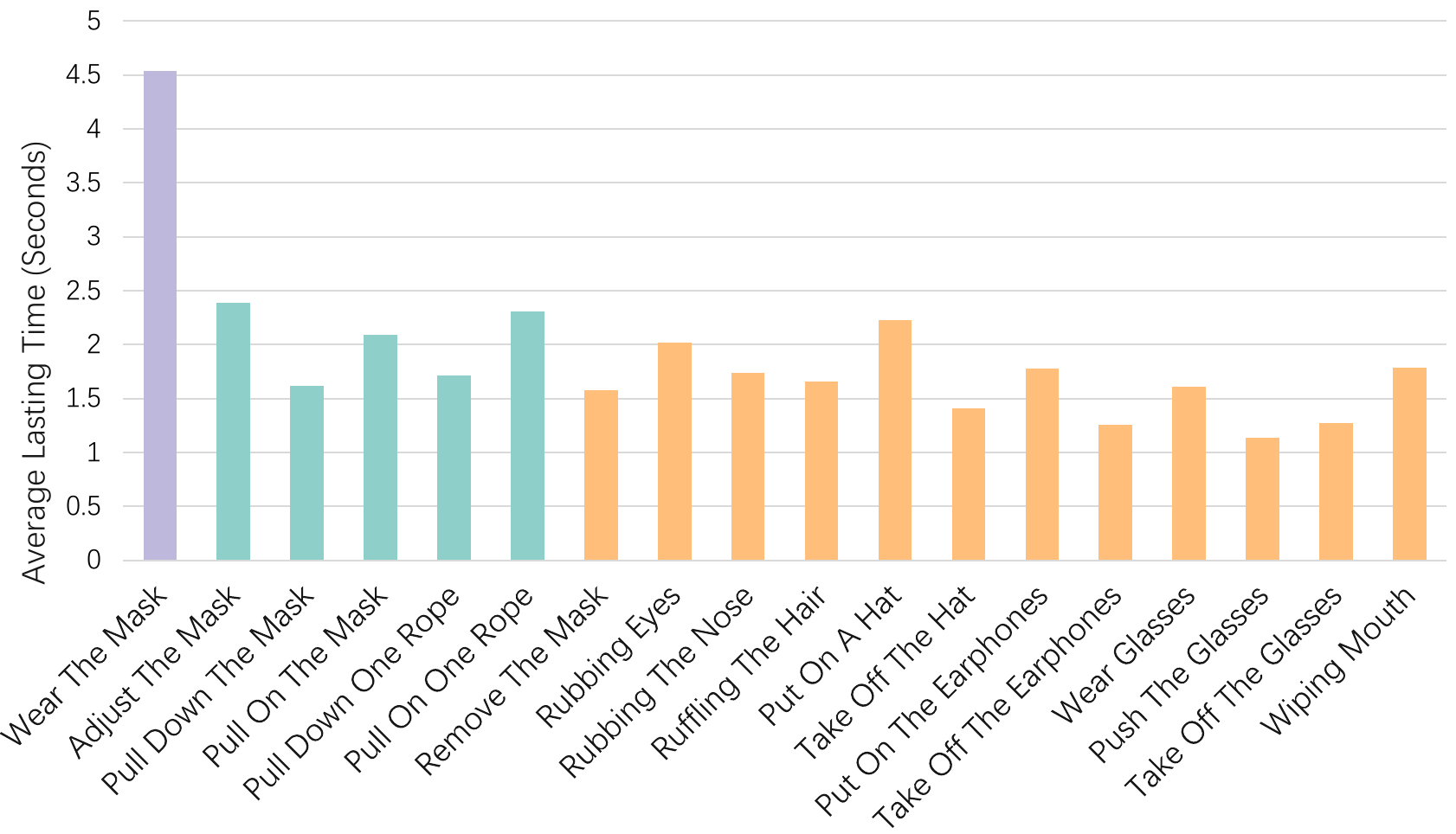}
    \setlength{\abovecaptionskip}{-10pt}
    \caption{Mask-related hand activities is diverse in time and activity. }
    \label{fig:diversity}
\end{figure}

We find MLP-Mixer~\cite{mlp-mixer} is a suitable algorithm to overcome our challenge. In MaskReminder, we first cut a time series into several clips, then apply a multi-layer perceptron~(MLP) on every individual clip, which aims to learn intra-clip~(local/short-range) features from the data of accelerators and gyroscopes. To learn inter-clip~(global/long-range) features, MLPs are applied to the feature space dimension by dimension across all clips. The intra-clip MLPs and inter-clip MLPs are applied cyclically for more discriminative features to estimate the mask-wearing status of users.

Other challenges arise from the diversity of hand movements. (1) Diversity in activity. MaskReminder estimates the mask-wearing status via the hand activities such as wearing a mask, removing the strap from behind the ears, and lowering the metal strap. However, hand movements that interact with the face or head e.g., rubbing eyes, wearing glasses, and wearing a hat, may mislead MaskReminder. To overcome this problem, we collect data of additional 11 activities, as shown in Fig.~\ref{fig:diversity}, and enable MaskReminder to resist the misleading of these activities in mask-wearing status estimation. (2) Diversity in time. Wearing a mask may last 4.5 seconds, while lowering the metal strap for drinking may last only 1.5 seconds. If MaskReminder makes the mask-wearing status estimation based on 4.5-second recordings, features of fast activities may be drowned, leading to a missing estimation. To solve this problem, we compute the average lasting time of all activities, as shown in Fig.~\ref{fig:diversity}, and set the estimation duration to 2.56 seconds (2.56$\times$50Hz=128 points). This parameter serves as the crop operation on the activities of wearing a mask, making MaskReminder detect mask-wearing even not see the whole procedure. Significantly, this parameter will introduce extra sampling points into estimation recordings of fast activities, enabling MaskReminder resilient to noise introduced by these extra points.

We recruit 20 participants to evaluate the performance of MaskReminder. We adopt MLP-Mixer to estimate mask-wearing status and achieved 89\% accuracy. When compared to ResNet~\cite{resnet} and SVM, MLP-Mixer outperformed.
Contributions of this paper can be summarized as three folds.

(1) We propose MaskReminder, a smartwatches-based prototype, to estimate whether the user wears a mask, for reminding those who may be involved in the COVID-19 virus transmission scenarios.

(2) We adapt MLP-Mixer and demonstrate its advances in learning local and global information from time-series data.

(3) We experiment on 20 participants and 8000+ data samples, which shows that MaskReminder can accurately estimate the mask-wearing status in both the user-dependent evaluation and the user-independent evaluation.

\subsection{System Workflow}\label{sec:systemoverview}

% MaskReminder is to remind users who may involve in COVID-19 virus transmission scenarios to wear a mask to protect themselves. 
MaskReminder intends to remind users who may be involved in COVID-19 virus transmission scenarios to wear a mask for self-protection. To build MaskReminder, we follow the standard supervised learning workflow, i.e., train MaskReminder with annotated data and then deploy it for use. In the training phase, we recruit volunteers to wear smartwatches to conduct activities shown in Fig.~\ref{fig:diversity}, collect recordings of accelerometers and gyroscopes, annotate these recordings with the corresponding category of activities, and train MaskReminder with paired recordings and annotations. In the use phase, MaskReminder reads the recordings of accelerometers and gyroscopes of smartwatches continuously and conducts mask-wearing status estimation over the recordings. MaskReminder will keep silent if it detects the user has worn a mask in a recent period. Otherwise, it will pop up a notification to alert the user to wear a mask. Fig.~\ref{fig:workflow} demonstrates the workflow.

% We envision MaskReminder works in people's daily life in several scenarios to prevent the transmission of the COVID-19 virus, e.g.,
% \begin{itemize}

% \item along with strength of GPS signals to automatically remind patients who suffer from allergic rhinitis or respiratory infections to wear a mask before going outside.

% \item along with map apps to remind users at public places such as bus stops, metro stations, malls, airports, etc.

% \item along with ride-hailing apps such as Uber, Lift, and DiDi to remind those who are going to take a car ride.

% \end{itemize}

% \subsection{Problem Formulation}\label{sec:problemformulation}
% The MaskReminder shall detect a series of actions that will occur in continuous mask wearing, such as wearing a mask, pulling down the mask to breathe, adjusting the metal clip of the mask and the lower edge of the mask, etc., to determine whether the user is wearing a mask, if not, the system will remind the user Wear a mask in time.

% Specifically, input the dataset $D$ =  $\{ A_i,G_i;Y_i \mid i \in 1,2,...,N \} $ into the system, where the $A$, $G$ represent acceleration data and gyroscope data, respectively,and $Y$ represent the action category. Then use the deep learning model $M$ to train data, and output action predictions, $Y^*$.

\section{Methods}~\label{sec:methond}

\subsection{Deep Network Framework}\label{sec:deepframwork}

We are to formulate the training of MaskReminder mentioned above. After data collection and labeling, we have paired recordings (from the accelerometers and gyroscopes) and annotations, denoted as $A$, $G$, and $y$ respectively. Given that we have N paired training data, our goal is to propose and train a model $\mathcal{W}$ that has minimize accumulative distances between its predictions and annotations. 
\begin{equation}\label{eq:goal}
    \mathcal{W} = \arg\min \sum_{i=1}^{N} distance(\mathcal{W}(A_i,G_i),~y_i)
\end{equation}
In MaskReminder, we utilize a MLP-Mixer variant, a deep network, as $\mathcal{W}$, described below.

\begin{figure}[t]
    \centering
    \includegraphics[width=1\linewidth]{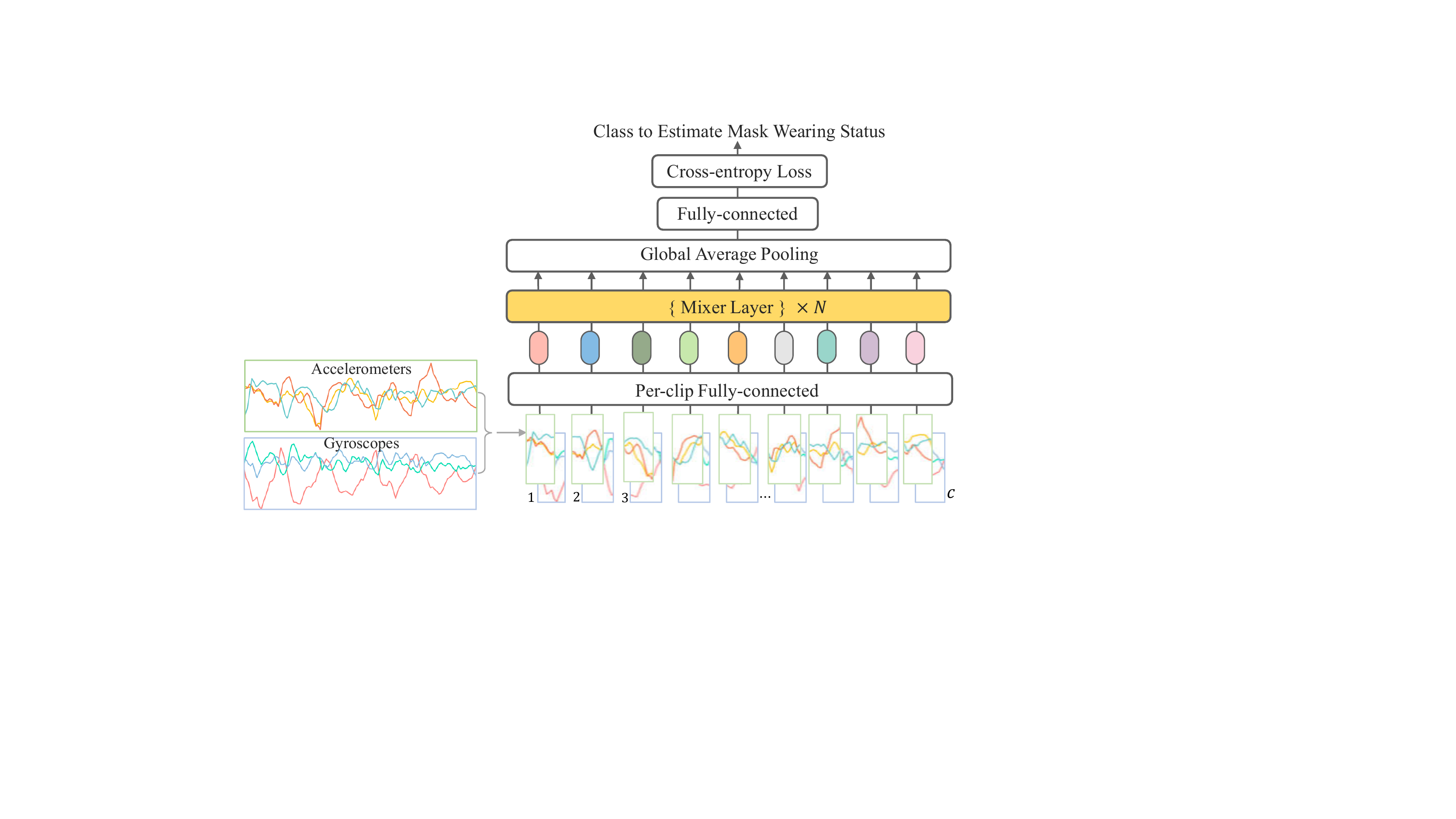}
    \setlength{\abovecaptionskip}{-10pt}
    \caption{MLP-Mixer. Raw recordings of accelerometers and gyroscopes of smartwatches are evenly divided, added, and flattened to be the inputs of MLP-Mixer. Per-clip fully-connected module is to convert the inputs from raw data space to hidden feature space. Then $N$ stacked Mixer layers are to mine the local and global features of the inputs. Further, MLP-Mixer outputs the activity category for the mask-wearing status estimation.}
    \label{fig:mlpmixer}
\end{figure}

\textbf{(1) Inputs.} Recall that accelemetors and gyroscopes measure physical values in three dimensions. As shown in Fig.~\ref{fig:mlpmixer}, inputs of MLP-Mixer, $A_i$ and $G_i$, are both with size of $L\times3$, where $L$ is for the length of $A_i$ and $G_i$. Referring ViT~\cite{vit}, we first cut $A_i$ and $G_i$ into $c$ clips evenly, leading to $a_i^j \in \mathbb{R}^{\frac{L}{c}\times 3}$ and $g_i^j \in \mathbb{R}^{\frac{L}{c}\times 3}$, where $a_i^j$ and $g_i^j$ are for the $j-$th clip of the $A_i$ and $G_i$, respectively. All input recordings are raw data, i.e., without any pre-processing.

\textbf{(2) Per-clip Fully-connected.} Each pair of $a_i^j$ and $g_i^j$ is concatenated along the 2nd dimension to be $x_i^j \in \mathbb{R}^{\frac{L}{c}\times 6}$. Then we flatten $x_i^j$ to be with size of $\mathbb{R}^{\frac{6L}{c}}$. The per-clip fully connected module is implemented via one fully connected layer without bias to maps $x_i^j \mapsto e_i^j\in \mathbb{R}^h$. This module works as a lightweight embedding function to convert inputs from raw data space to hidden feature space.  

\textbf{(3) Mixer Layer.} The Mixer layer is the key component of MLP-Mixer, which is capable of efficiently mining the intra-clip~(local) and inter-clip~(global) representation of inputs for hand activity recognition, described in Sec.~\ref{sec:Mixerlayer}.

\textbf{(4) Classification Head.} The head consists of a global average pooling and a fully connected layer. Given that the input size of global average pooling is $c\times h^*$, where $h^*$ is for the output dimension of Mixer layers on each $e_i^j$. The global average pooling conducts the pooling operation across the $clip$ dimension, which further mixes features from all clips to gain a global view for the hand activity classification and mask-wearing status estimation.

\begin{figure*}[t]
\label{sec:mixerlayer}
    \centering
    \includegraphics[width=1\linewidth]{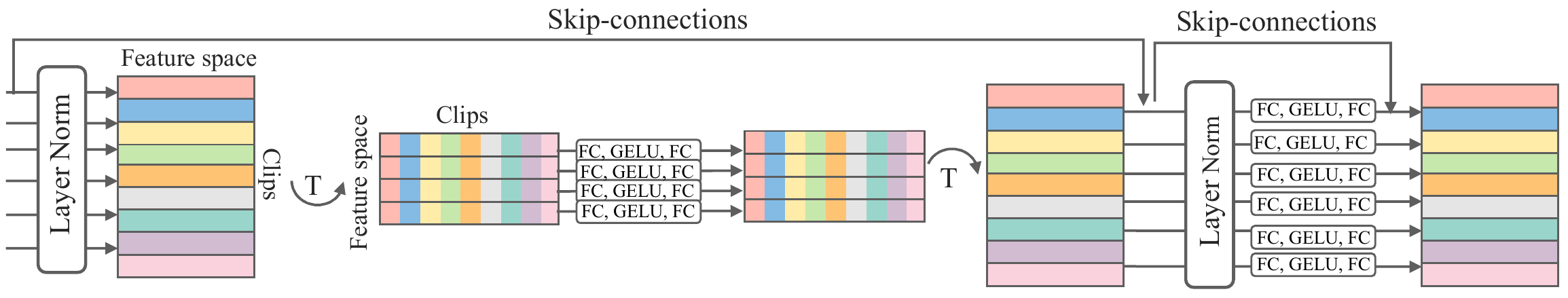}
    \setlength{\abovecaptionskip}{-10pt}
    \caption{Mixer Layer. Mixer layer first conducts \{FC, GELU, FC \} over the row of transposed clips, learning global representation across clips. Then it conducts a 2nd \{FC, GELU, FC \} over the row of clips, learning local representation in the clip. With alternate representation learning from global and local views, Mixer layer outputs good representation to estimate the status of mask-wearing.} 
    \label{fig:mixerlayer}
\end{figure*}

\subsection{Mixer Layer}\label{sec:Mixerlayer}

The Mixer layer serves as the key representation learner in the middle of MLP-Mixer, as shown in Fig.~\ref{fig:mlpmixer}. We further illustrate its details in Fig.~\ref{fig:mixerlayer}. Recall that the per-clip fully-connected module of MLP-Mixer maps each divided clip $x_i^j \mapsto e_i^j\in \mathbb{R}^h$, thus $c$ clips lead to the inputs of Mixer layer $e_i\in \mathbb{R}^{c\times h}$.
The Mixer layer conducts \{FC~(fully-connected), GELU~(Gaussian Error Linear Unit)~\cite{hendrycks2016gaussian}, FC\}, across the columns and rows of $e_i$ respectively to learn the global and local representation of inputs respectively.

\textbf{(1) Inter-clip mixing for global representation.} As shown in Fig.~\ref{fig:mixerlayer}, we first apply Layer Normalization~\cite{ba2016layer} on $e_i$ and get LN$(e_i)$. Then we transpose LN$(e_i)$ to LN$^T(e_i) \in \mathbb{R}^{h\times c}$, and conduct \{FC, GELU, FC\} operations along every row of LN$^T(e_i)$, i.e., every column of LN$(e_i)$. For elements of every column of LN$(e_i)$ are comprised of features from all clips in the whole recordings, this sequence of operations is to mix feature of inter-clips, learning global representation of recordings. We represent these operations as follows.
\begin{equation}\label{eq:global}
(u_i)_{*, j} = (e_i)_{*,j}+\mathbf{W}_2\,\sigma\bigl( \mathbf{W}_1\, \text{LN}({e_i})_{*,j} \bigr), \text{for } j=1\ldots h   
\end{equation}
where $\sigma$ is for GELU, a non-linear activation function; $\mathbf{W}_1$ is for parameters of the fully-connected layer; $\mathbf{W}_2$ is for parameters of the fully-connected layer; $(u_i)$ is the output.

\textbf{(2) Intra-clip mixing for local representation.} As shown in Fig.~\ref{fig:mixerlayer}, once we have $u_i$, we first transpose it to $u_i^T \in \mathbb{R}^{c\times h}$. Then we apply Layer Normalization on $u_i^T$ and get LN$(u_i^T)$. , and conduct \{FC, GELU, FC\} operations along every row of LN$(u_i^T)$. For elements of every row of LN$(u_i)^T$ are comprised of features from one clip, this sequence of operations is to mix features of intra-clip, learning local representation of every clip. We represent these operations as follows.
\begin{equation}\label{eq:local}
(v_i)_{k, *} = (u_i)_{k,*}+\mathbf{W}_4\,\sigma\bigl( \mathbf{W}_3\, \text{LN}({u_i})_{k,*} \bigr), \text{for} k=1\ldots c   
\end{equation}
where $\sigma$ is for GELU, a non-linear activation function; $\mathbf{W}_3$ is for parameters of the third fully-connected layer; $\mathbf{W}_4$ is for parameters of the fourth fully-connected layer; $(v_i)$ is the output.

One Mixer layer consists of operations of Eq.~\ref{eq:global} and Eq.~\ref{eq:local}, which conducts intra-clip mixing and inter-clip mixing for global and local representation learning. As shown in Fig.~\ref{fig:mixerlayer}, we can stack multiple Mixer layers for a more powerful representation to estimate the mask-wearing status with the CrossEntropy loss function. We show the pseudocode of MLP-Mixer with Pytorch-like style in Algorithm~\ref{alg:code}.

\subsection{Training Details}\label{sec:trainingdetails}

We implement MLP-Mixer with Pytorch 1.10.2 and train it with a single RTX3090. We train the network to a maximum of 400 epochs, and set up an early stop mechanism. The patience on monitoring training loss is 40. The initial learning rate is 0.0005, which decays by 0.5 every 40 epochs. We use AdamW~($\beta_1 = 0.9, \beta_2 = 0.999$) to optimize MLP-Mixer. 

%##################################################################################################
\begin{algorithm}[t]
\caption{Pseudocode of MLP-Mixer in Pytorch-like.}
\label{alg:code}
\algcomment{\fontsize{7.2pt}{0em}\selectfont 
%\vspace{-1.em}
}
\definecolor{codeblue}{rgb}{0.25,0.5,0.5}
\lstset{
  backgroundcolor=\color{white},
  basicstyle=\fontsize{7.2pt}{7.2pt}\ttfamily\selectfont,
  columns=fullflexible,
  breaklines=true,
  captionpos=b,
  commentstyle=\fontsize{7.2pt}{7.2pt}\color{codeblue},
  keywordstyle=\fontsize{7.2pt}{7.2pt},
%  frame=tb,
}
\begin{lstlisting}[language=python]
# Linear: fully connected layer
# GELU: GELU activation function
# GAP: global average pooling

mlpmixer.params # initialize
mixer_layer_num = n
for x in loader: # load a minibatch x with N samples
    #divide x to c clips, Nx6xL -> Nx6xcxL/c
    [x1,x2,x3,..,xc] = x
    
    # Per-clip fully-connected layer
    e = Linear([x1,x2,...,xc]) #e: Nxcxh
    
    # mixer layers
    for mixer_layer_num:
        # inter-clip mixing for global representation
        # u: Nxhxc
        e = transpose(e)
        u = e + Linear(GELU(Linear(LayerNorm(e))))
         
        # intra-clip mixing for local representation
        # v: Nxcxh
        u = transpose(u)
        v = u + Linear(GELU(Linear(LayerNorm(u))))
         
        mixer_layer_num -= 1
    
    # classification head
    v = GAP(v) # v: Nxh
    loss = CrossEntropyLoss(Linear(v),labels)
    
    #Adam update MLP-Mixer
    loss.backward()
    update(mlpmixer.params)
\end{lstlisting}
\end{algorithm}

\section{Evaluation}~\label{sec:experiment}

\subsection{Data Acquisition}~\label{subsec:dataco}

We recruit 20 volunteers, 7 men and 13 women, to conduct actions shown in Fig.~\ref{fig:diversity}. These actions are divided into three categories, i.e., (1) Wearing a mask. (2) Mask-wearing related actions, e.g., adjusting the mask to ensure no gap at the face and nose, pulling down the mask for breath. (3) Actions that may mislead mask-wearing status estimation, e.g., rubbing eyes or nose, putting on or taking off a hat or earphones. Each volunteer repeats each action 20 times (10 times when smartwatch on the right hand, 10 times on the left hand).  We use a Samsung Gear Sport smartwatch to record readings of accelerometers and gyroscopes with 50Hz, and corresponding timestamps. Meanwhile, we use a camera to record video streams and corresponding timestamps. Timestamps from the smartwatch and camera are used to synchronize IMU readings and videos. Then we replay and watch video streams to label the start time and end time of each action repeated, with which we segment IMU readings. In all, we have a dataset with 7200 segments of IMU readings (20 volunteers$\times$18 actions$\times$2 hands$\times$10 repeats). 

For these segments with different lengths, we normalize their length to 128 as follows. (1) If the length of one segment $L$ is less than 50, we discard this segment from the dataset. (2) If $L \in [50,128]$, we append $128-L$ zeros to the segments. (3) If $L > 128$, we cut the segment into multiple clips by every 128 sampling points without overlapping, and conduct (1) and (2) over the last clip. After this normalization operation, the dataset is with 8039 segments.

\begin{table}[t]
\centering
\caption{Results of MLP-Mixer, SVM and ResNet. `Layers' is the number of Mixer layer shown in Fig.~\ref{fig:mixerlayer}. `Dim.' is for the output dimension of the Per-clip Fully-connected layer, described in Sec.~\ref{sec:deepframwork}. `Len.' is for the clip length of one segment divided into several clips, shown in Fig.~\ref{fig:mlpmixer}.}
\label{tab:Comparison}
\renewcommand{\arraystretch}{1.2}
\setlength\tabcolsep{2pt}
\begin{tabular}{ccccccc}
\hline
Model & Layers & Dim.  & Len.  & Parameters(M) & FLOPs(M) & Accuracy \\\hline
SVM~(statistic)  & $\backslash$   & $\backslash$  & $\backslash$ & $\backslash$   & $\backslash$   & 0.68   \\
SVM~(simple)  & $\backslash$   & $\backslash$  & $\backslash$  & $\backslash$   & $\backslash$   & 0.74  \\\hline
ResNet/18  & $\backslash$   & $\backslash$  & $\backslash$  & 3.85   & 22.15   & 0.80    \\
ResNet/34  & $\backslash$   & $\backslash$  & $\backslash$  & 7.22   & 45.12   & 0.72     \\
ResNet/50  & $\backslash$   & $\backslash$  & $\backslash$  & 15.96   & 101   & 0.81      \\
ResNet/101  & $\backslash$   & $\backslash$  & $\backslash$  & 28.26   & 200   & 0.77     \\\hline
Mixer/ES/32  & 2      & 128   & 32  & 0.29   & 1.19   &0.81     \\
Mixer/ES/16 & 2  & 128   & 16    & 0.28   & 2.35   &0.85      \\
Mixer/ES/8  & 2  & 128   & 8  & 0.28   & 4.87   & 0.88        \\
Mixer/MS/32 & 4  & 256   & 32  & 2.16   & 8.76   &0.85   \\
Mixer/MS/16   & 4  & 256   & 16 & 2.13  & 17.59  &0.87       \\
\textbf{Mixer/MS/8}  & 4  & 256   & 8   & 2.13   & 36.03   & $\textbf{0.89}$    \\
Mixer/S/32  & 8  & 512   & 32   & 16.91   & 68.2   &0.85    \\
Mixer/S/16  & 8  & 512   & 16  & 16.87   & 137   &0.87      \\
Mixer/S/8  & 8  & 512   & 8  & 16.85   & 278   & 0.88   \\
\hline
\end{tabular}
\end{table}

\begin{figure}[t]  
    \centering
    \includegraphics[width=1\linewidth]{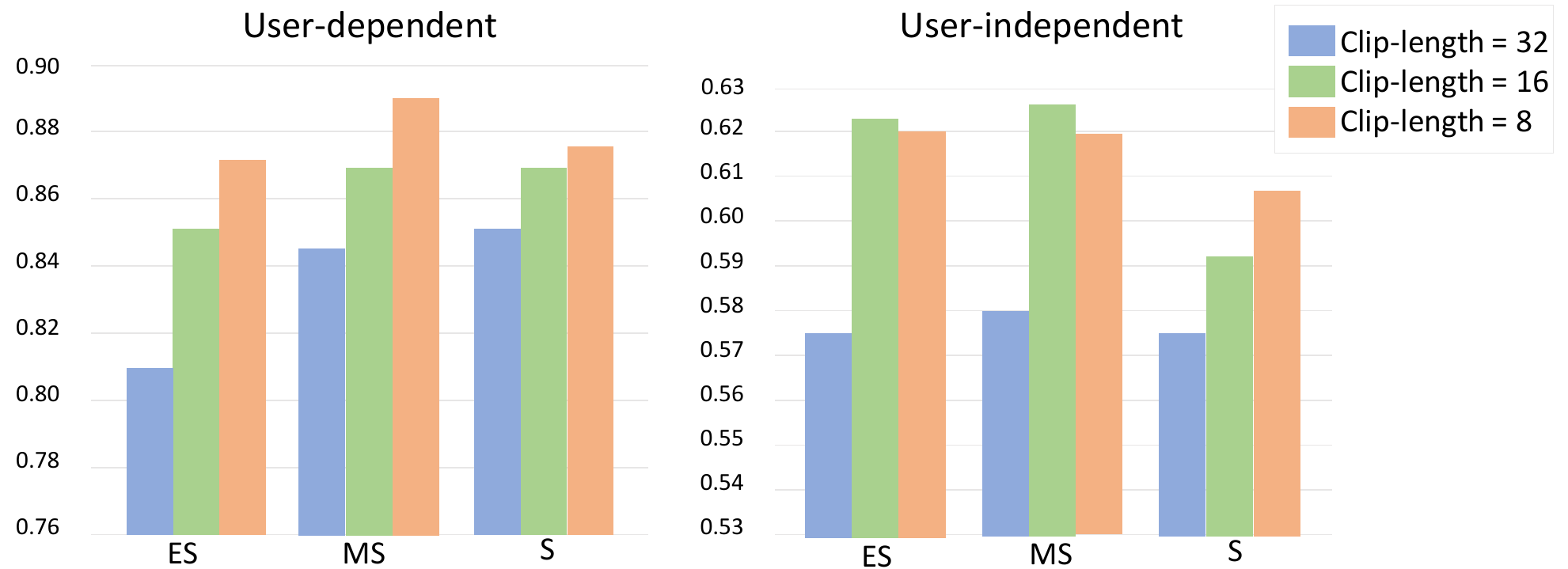}
    \setlength{\abovecaptionskip}{-10pt}
    \caption{Comparison of different scales and clip lengths over MLP-Mixer. All large models face the problem of over-fitting. }
\label{fig:ablation} 
\end{figure}

\begin{figure}[t]
    \centering
    \includegraphics[width=0.98\linewidth]{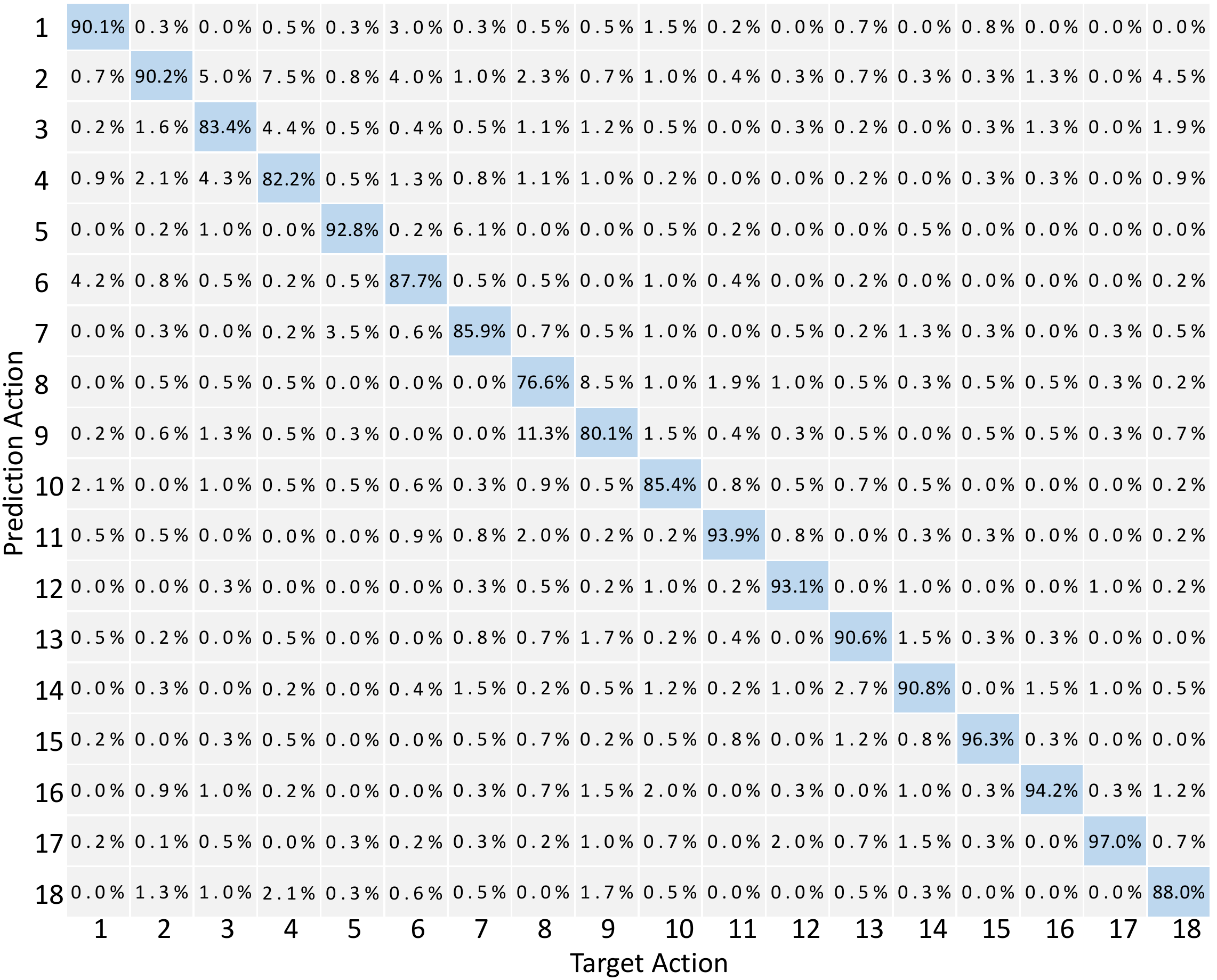}
    \setlength{\abovecaptionskip}{-10pt}
    \caption{Confusion matrix on 18 activities. MaskReminder accurately distinguishes mask-wearing-related actions (1-6) and interfering actions (7-18). }
    \label{fig:action}
\end{figure}

\subsection{Performance}~\label{subsec:results}

We evaluate MaskReminder with the collected dataset in user-dependent manner and user-independent manner as below.

\textbf{(1) User-dependent.} For segments of each action of each volunteer, we divide them into 5 groups according to the order of conducted time, denoted as $G_1$, $G_2$, $G_3$, $G_4$, $G_5$. We first apply $G_2-G_5$ to train the MLP-Mixer and test the trained MLP-mixer with $G_1$. Then we use $G_3-G_5$ and $G_1$ to train and $G_2$ to test. We apply this leave-one-group-out evaluation across all groups, obtain 5 trained models, and report the performance next. 

$\bullet$ Table~\ref{tab:Comparison} shows Mixer/MS/8 achieves the action recognition accuracy of 0.89, outperforming all other MLP-Mixers and demonstrating that MaskReminder performs well to estimate mask-wearing status.

$\bullet$ Table~\ref{tab:Comparison} indicates an over-fitting phenomenon happens. That is, for example, the largest model (MLP-Mixer/S/8 with 278 MFLOPs) is inferior to a smaller one, e.g., MLP-Mixer/MS/8 with 36.03 MFLOPs. Further, we visualize the results into 3 groups according to parameters in Fig.~\ref{fig:ablation}, i.e., ES, MS, and S, which shows models with more FLOPs perform better when with similar parameters. 

$\bullet$ Table~\ref{tab:Comparison} also shows that MLP-Mixers largely outperform a traditional method, i.e., Support Vector Machine~(SVM), and a modern method, i.e., ResNet. SVM (statistic) is computed with statistic features of IMU segments such as the average, skewness, kurtosis, entropy, etc., as in~\cite{thomaz2015practical}. SVM (simple) is computed with raw IMU readings as features. ResNets are computed by replacing the 2D convolution and 2D pooling with 1D convolution and 1D pooling that swiping along the time dimension of IMU segments, inspired by~\cite{wang2019joint}.

$\bullet$ Fig.~\ref{fig:action} shows the confusion matrix of the action recognition, which shows that MaskReminder can accurately~($\approx$0.90) recognize all actions. The most errors occur in the prediction between two very similar actions, i.e., the 8th action~(rubbing eyes) and the 9th action~~(rubbing the nose). 

% through the Fig.~\ref{fig:action}
%  Of all the 18 actions captured ($1$ includes adjustments while wearing the mask and adjustments during wearing the mask),

\begin{figure}[t]
    \centering
    \includegraphics[width=1\linewidth]{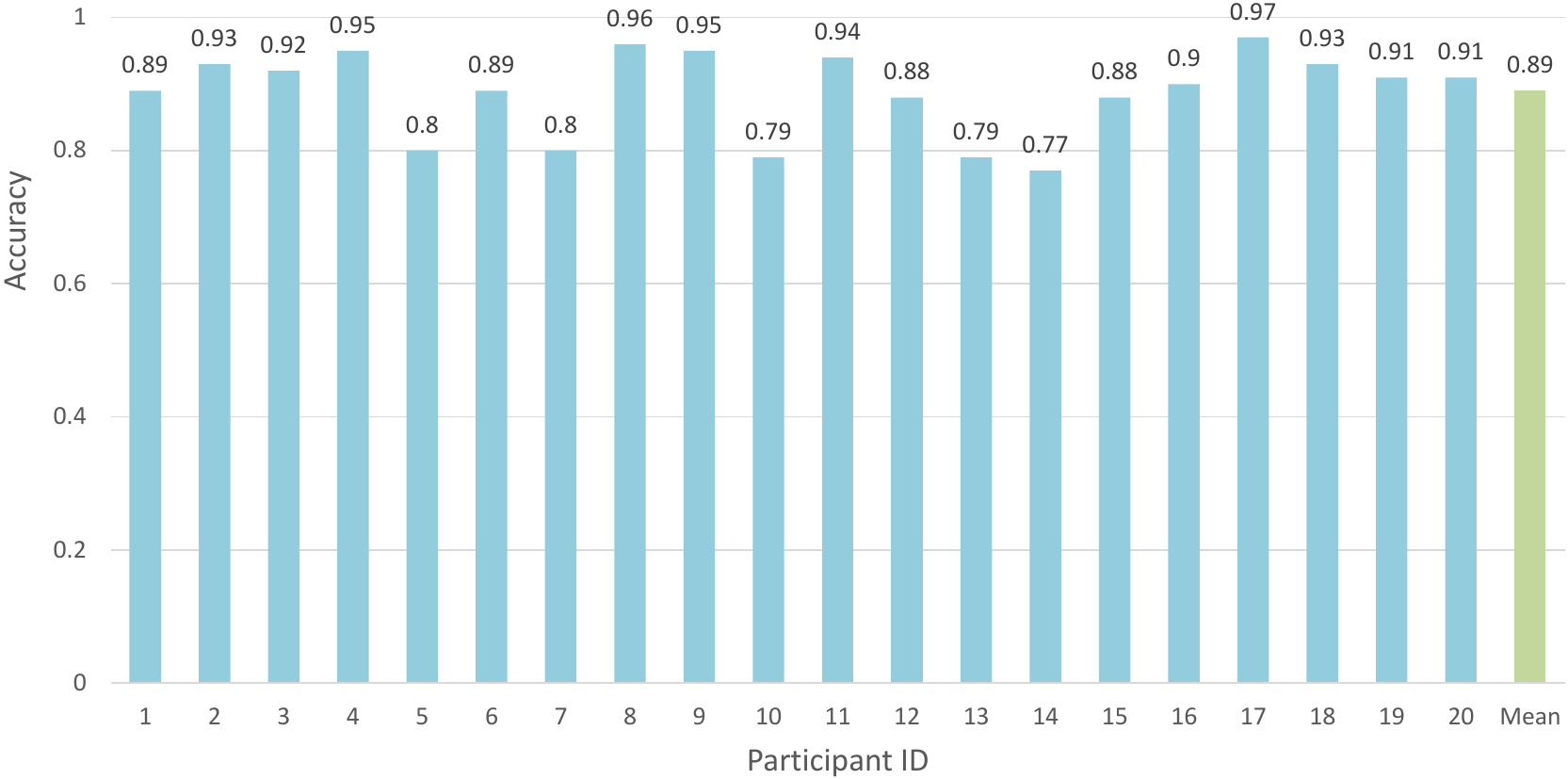}
    \setlength{\abovecaptionskip}{-10pt}
    \setlength{\belowcaptionskip}{-10pt}
    \caption{MaskReminder with MLP-Mixer/MS/8 performs well over all the subjects. The mean accuracy is 0.89. }
    \label{fig:normalAccuracy}
\end{figure}

$\bullet$ Recall that we recruit 20 volunteers to evaluate MaskReminder. We compute the action recognition accuracy of each volunteer. Fig~\ref{fig:normalAccuracy} shows that the average accuracy of the recognition of the wearing state of the mask is 0.89, and the highest accuracy rate can reach 0.97.

$\bullet$ Recall that volunteers wear the smartwatch on their right hand and left hand respectively. We further evaluate the trained models on segments of the right hand and left hand respectively. The mean accuracy of the right hand and left hand is 0.91 and 0.87, respectively. That is why, for most volunteers, the right hand is their dominant hand, IMU readings from the right hand can provide more movement-related information.  

% \begin{equation}
%     A_p = \frac{1}{5}\sum_{i=1}^{5} \sum_{j=1}^{G_i^p}\frac{\left\| W(X_j^G_i^p),y_j^G_i^p) \right\|}{ G_i^p }
% \end{equation}

% \text{(1)} First, the normal strategy: we select two of the ten repeated instances of each subject as the test set, and the other eight instances as the training set, to check MaskReminder's capacity to learn the subject's action patterns and predict the class of the subject's future action. Each action has ten instances, from which we select two at a time and repeat five times without overlap in total, averaging the results, Fig~\ref{fig:normalAccuracy} shows that the average accuracy of the recognition of the wearing state of the mask is 89\%, and the highest accuracy rate can reach 97\%. From the action recognition perspective, we show the results as in the Fig~\ref{fig:action}, where $0-17$ represents the 18 action categories. We can observe that 0th~(Rubbing eyes) and 1th~(Rubbing noses) exists some similarities.

% And considering the different hand habits of subjects, the test set we obtained under the normal strategy was separated by subject’s hand: left and right. But we train the model using both left-hand and right-hand data. In this way, we can train one model to fit different hand habits of subjects. As the result, the mean accuracy of the left is 86.5\%, and the right is 91.2\%, which suggests the model can fit different hand habits just fine and can learn from one subject’s historical data to predict the particular subject’s future action.
 
% \begin{table}[t]
% \caption{Accuracy over left/right test set.}
% \centering
% \renewcommand{\arraystretch}{1.2}
% \begin{tabular}{cccc}
% \hline
% Test data of         & Left Hand & Right Hand \\ \hline
% Accuracy    & 0.87 & 0.91  \\ \hline
% \end{tabular}
% \label{tab:left-right-accuracy}
% \end{table}

\begin{figure*}[t]
    \centering
    \includegraphics[width=1\linewidth]{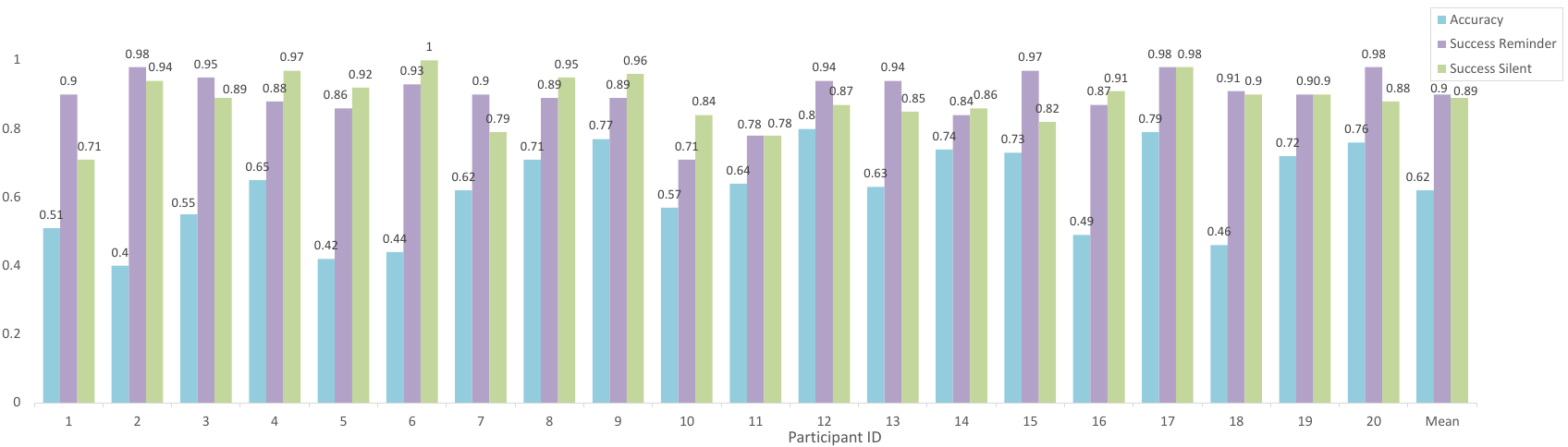}
    \setlength{\abovecaptionskip}{-15pt}
    \caption{MaskReminder results of the leave-one-subject-out manner. The mean accuracy, rate of success reminder, and rate of success silence are 0.62, 0.90, and 0.89, respectively.}
    \label{fig:user}
\end{figure*} 

\textbf{(2) User-independent.} In this manner, we train MLP-Mixer with segments of 19 out of 20 volunteers, and test the trained model with the segments of the remaining volunteer. We apply this leave-one-user-out evaluation across all volunteers, obtain 20 trained models, and show the results in Fig.~\ref{fig:user}. 

As Fig.~\ref{fig:user} shows, when we train MLP-Mixer/MS/8 with segments of the 2nd to 20th volunteers and evaluate the train model with segments of the 1st volunteer, the accuracy is 0.52, not quite satisfactory. However, it is worth mentioning that MaskReminder is designed to report whether the user is wearing a mask. Thus, if actions that indicate the mask-wearing status, e.g., 7th to 18th in Fig.~\ref{fig:diversity}, are classified as actions of 7th to 18th (even if not correctly classified as the target action), MaskReminder can still remind successfully.
We call this success reminder. Similarly, if MaskReminder classifies one segment of 1st to 6th in Fig.~\ref{fig:diversity} as the 1st to 6th, it can keep successfully silent. This binary classification largely reduces the requirement of MaskReminder. Overall, the mean accuracy, rate of success reminder, and rate of success silence are 0.62, 0.90, and 0.89, respectively, indicating MaskReminder is still a promising approach for estimating the mask-wearing status in the user-independent setting for unseen users.

\section{Related Work}~\label{sec:related_work}

\subsection{Mask Wearing Detection}
Some works are proposed to detect whether people are wearing masks with the algorithms on images or videos~\cite{kong2021real,das2020covid,yang2022mask}, but these works can only be applied at places where cameras are deployed, thus cannot give users a timely reminder to help them wear masks to block the spread of the virus.
% Besides, ~\cite{Warning} has developed a mask recognition alarm system based on smart glasses, which recognizes people who are not wearing masks through images and then alerts disabled users.
In~\cite{maskstudy}, self-compliance with personal management regulations and conscious wearing of masks are more effective in preventing the spread of infectious diseases than from outside supervision. Therefore, \cite{Wristband} proposed a method to prevent infectious diseases by using a wristband with an inertial measurement unit ~(IMU), to detect whether the person is wearing a mask. However, since wristband devices are rarely used and are not aesthetically pleasing or convenient. So we choose the smartwatch for the MaskReminder system.

\subsection{Smartwatches}
Smartwatches already perform well in hand movement tracking~\cite{armtracking}. In this context, a series of studies of human activity recognition have emerged. The authors in ~\cite{vu2018smartwatch} utilize machine learning approaches to recognize and track people's hand movements. ~\cite{hou2019signspeaker} proposes a prototype system to recognize American sign language using a long short-term memory model. Smokewatch~\cite{smokewatch}is a smartwatch application that uses sensors to recognize hand movements and help smokers willing to quit.  In ~\cite{svmsmartwatch}, the author extracts the magnitude of hand movements from the smartwatch data, through a support vector machine ~(SVM) classifier that detects drivers' drowsiness.  UWash~\cite{handwashing} can process sample-wise handwashing gesture classification using a unified U-Net variant.

\section{Conclusion}~\label{sec:conclusion}

In this paper, we present a smartwatch-based mask reminder system - MaskReminder, which can detect user's hand movements via the built-in IMU sensors of smartwatches and estimate the mask-wearing status. MaskReminder adopts MLP-Mixer models to learn the local and global information from time-serial IMU readings. Extensive experimental results from over 20 participants demonstrate that MaskReminder can accurately estimate the mask-wearing status in both the user-dependent evaluation and the user-independent evaluation. We envision MaskReminder expanding the functionality of smartwatches for preventing virus transmission in the current COVID-19 pandemic in people's daily life.

{\small
\bibliographystyle{./bibliography/IEEEtran}
\bibliography{./bibliography/reference}
}

\end{document}